\documentclass[twocolumn,showpacs,pra,footinbib,superscriptaddress]{revtex4}
\usepackage{graphicx}
\usepackage{epstopdf}
\usepackage{amsfonts}
\usepackage{amsmath}
\usepackage{times}
\usepackage{cancel}
\usepackage{soul}
\newcommand{\ket}[1]{| #1 \rangle}

\newcommand{\ptrace}[2]{\mbox{$\mathrm{Tr}_{#1}$}(#2)}

\newcommand{\bra}[1]{\langle #1 |}

\begin{document}

\title{Quantum Interactions with Closed Timelike Curves and Superluminal Signaling}

\author{Jeffrey Bub}
\affiliation{Philosophy Department, University of Maryland, College Park, MD 20742, USA}
\affiliation{Institute for Physical Science and Technology, University of Maryland, College Park, MD 20742, USA}
\author{Allen Stairs}
\affiliation{Philosophy Department, University of Maryland, College Park, MD 20742, USA}
\date{\today}

\begin{abstract}
There is now a significant body of results on quantum interactions with closed timelike curves (CTCs) in the quantum information literature, for both the Deutsch model of CTC interactions (D-CTCs) and the projective model (P-CTCs). As a consequence, there is a prima facie argument exploiting entanglement that CTC interactions would enable superluminal and, indeed, effectively instantaneous signaling.  In cases of spacelike separation between the sender of a signal and the receiver, whether a receiver measures the local part of an entangled state  or a disentangled state to access the signal can depend on the reference frame. We propose a  consistency condition that gives priority to either an entangled perspective or a disentangled perspective in spacelike separated scenarios. For D-CTC interactions, the consistency condition gives priority to frames of reference in which the state is disentangled, while for P-CTC interactions the condition selects the entangled state. Using the consistency condition, we show that there is a procedure that allows Alice to signal to Bob \emph{in the past} via relayed superluminal communications between spacelike separated Alice and Clio, and spacelike separated Clio and Bob. This opens the door to time travel paradoxes in the classical domain. Ralph  \cite{Ralph2011} first pointed this out for P-CTCs, but we show that Ralph's procedure for a `radio to the past' is flawed. Since both D-CTCs and P-CTCs allow classical information to be sent around a spacetime loop, it follows from a result by Aaronson and Watrous \cite{Aaronson+2009} for CTC-enhanced \emph{classical} computation that a quantum computer with access to P-CTCs would have the power of PSPACE, equivalent to a D-CTC-enhanced quantum computer.
\end{abstract}


\maketitle

\section{Introduction}
\label{intro}

It is well-known that General Relativity permits the existence of closed time-like curves (CTCs). There is now a significant body of results in the quantum information literature on quantum interactions with CTCs, offering a  possible window on what we might expect in a future theory of quantum gravity. 

In classical theories, CTCs are a potential source of paradoxes. The `Grandfather Paradox' raises the worry that someone might travel into the past and kill her grandfather before he has any children; the `Unproved Theorem' paradox points out that if there are CTCs, then it might be possible to take a published proof of a theorem into the past and present it to someone, who then uses it to produce the very manuscript that leads to the theorem's publication.  David Deutsch \cite{Deutsch1991} offered a quantum mechanical account of CTCs that was intended to rule out such paradoxes from the outset. On the Deutsch model, a quantum system traversing a closed timelike curve (the CTC system) must satisfy a fixed point consistency condition. If we think of the CTC system as entering the future mouth of a `wormhole' in spacetime in a state $\rho_{CTC}$  and emerging from the past mouth in the state $\rho'_{CTC}$, Deutsch requires that $\rho'_{CTC} = \rho_{CTC}$. This restriction holds even if the CTC system interacts with a causality-respecting (CR) system; interactions between CR and CTC systems satisfy the condition:
\begin{equation}
\rho_{CTC} = \ptrace{CR}{U(\rho_{CTC}\otimes\rho_{CR} )U^{\dagger}}
\label{D-consis}
\end{equation}
It follows that the time-traveller on the CTC can't enter the loop in Grandpa's future, travel to the past and prevent her own birth by killing Grandpa. 
 
CTCs satisfying Deutsch's condition are referred to as D-CTCs. Since the interaction is a completely positive map with at least one fixed point, a suitable state $\rho_{CTC}$ always exists. If more than one state satisfies the condition, Deutsch imposes a maximum entropy condition to guarantee uniqueness. Requiring that $\rho_{CTC}$ be the same both entering and exiting the wormhole appears to stave off the Grandfather paradox, though we will argue later that this appearance is illusory. In any case, since $\rho_{CTC}$ depends both on $\rho_{CR}$ and itself, the resulting evolution is non-linear. As we will see, this can produce surprising behavior.

Bennett and Schumacher \cite{Bennett+2002} and, independently, Svetlichny \cite{Svetlichny2009}, offered an alternative model of CTCs, which was further developed by Lloyd \emph{et al.} \cite{Lloyd+2011a,Lloyd+2011b,Lloyd+2011c}. This account simulates CTC interactions by quantum teleportation  combined with postselection---hence, P-CTCs. To see the underlying intuition, consider a simple teleportation experiment. Clio has a qubit in state $|\psi\rangle_C$; Alice and Bob share a pair in the Bell state $\ket{\phi_{+}}=\frac{1}{\sqrt{2}}(|0\rangle_{A}|0\rangle_{B} + |1\rangle_{A}|1\rangle_{B})$. Alice measures qubits $A$ and $C$ in the Bell basis and communicates the result to Bob, who applies an appropriate unitary depending on which of the four outcomes occurred. Should the outcome of the Bell measurement correspond to the original Bell state $\ket{\phi_{+}}$, however, there is nothing Bob needs to do. Lloyd \emph{et al.} offer this comment:
\begin{quote}
In this case, Bob possesses the unknown state even before Alice implements the teleportation. Causality is not violated because Bob cannot foresee Alice's measurement result, which is completely random. But, if we could pick out only the proper result, the resulting `projective' teleportation would allow us to travel along spacelike intervals, to escape from black holes, or to travel in time. We call this mechanism a \mbox{projective} or postselected CTC, or P-CTC.
\end{quote}

Thus, P-CTCs model CTCs as cases of teleportation in which Nature, as it were, picks out the projection onto the appropriate entangled state. This  induces a different nonlinear evolution in the state of the CR system and can be interpreted as creating a quantum channel to the past.   Here the idea is that paradox is avoided because anything that could happen in a P-CTC interaction does happen with some non-zero probability in an ordinary quantum teleportation circuit. As Lloyd \emph{et al.} see it, this approach has many advantages, one of which is that it leads to testable predictions. What would happen in each case if we were in possession of a genuine P-CTC happens (in our simple case) for one quarter of the equally probably outcomes of the Bell measurement. This means that teleportation experiments combined with conventional post-selection can be used to test and illustrate predictions about CTCs.

With respect to the Deutsch model, Brun \emph{et al.} have shown that quantum information processors with access to D-CTCs could perfectly distinguish non-orthogonal quantum states \cite{Brun+2009} and also  clone any quantum state \cite{Brun+2013}.  Following previous work on the power of D-CTC-enhanced quantum computation by Bacon \cite{Bacon2004}, Brun \cite{Brun2003}, and Aaronson \cite{Aaronson2005a}, Aaronson and Watrous \cite{Aaronson+2009} proved that quantum computers with access to D-CTCs would have the power of the complexity class PSPACE, consisting of all problems solvable by a classical computer using a polynomial amount of memory---precisely the same as the power of classical computers with access to classical CTCs.  

For P-CTCs, Brun and Wilde \cite{Brun+2012} have  shown that suitable P-CTC interactions would enable an agent to `force' any particular outcome in a quantum measurement, where the outcome  would otherwise be an intrinsically random event. Following Aaronson's proof \cite{Aaronson2005b} that the complexity class PostBQP (post-selected bounded-error quantum polynomial-time) is equivalent to PP, Lloyd \emph{et al.} \cite{Lloyd+2011b} observed that  quantum computers with access to P-CTCs could compute any problem in PP in polynomial time. The complexity class PP is the class of decision problems that a probabilistic Turing machine, running in polynomial time, accepts with  probability at least $1/2$ if and only if the answer is `yes.' PP includes NP and is included in PSPACE (although it is still an open question whether the inclusions are strict). 

From the Brun \emph{et al.} results, there is a prima facie argument exploiting entanglement that both models of CTC interactions would enable superluminal and, indeed, effectively   instantaneous signaling, without anything physical passing between the source of the signal and the receiver.\endnote{For an opposing view, see C.H. Bennett, S. Leung, G. Smith, and J.A. Smolin, Phys. Rev. Lett. \textbf{103}, 170502 (2009), and the rebuttal by E.G. Cavalcanti and N.C. Menicucci, arXiv:1004.1219, and by Cavalcanti, Menicucci, and Pienaar \cite{Cavalcanti+2012}. The Bennett \emph{et al} paper raises issues about nonlinear extensions of quantum mechanics that we have set aside for the purposes of this paper and hope to address elsewhere.}  If the sender and receiver are spacelike separated, whether a receiver accesses the signal by measuring an entangled state  or a disentangled state can depend on the reference frame. We propose a  consistency condition that gives priority to either an entangled perspective or a disentangled perspective in spacelike separated scenarios. For D-CTC interactions, the consistency condition gives priority to frames of reference in which the state is disentangled, while for P-CTC interactions the condition selects frames in which the state is entangled.

It is possible for Alice and Clio to be spacelike separated, and Clio and Bob to be spacelike separated, so that Bob is in Alice's causal past.  Using the consistency condition, we show that there is a procedure that allows Alice to signal to Bob \emph{in the past} via relayed superluminal communications between Alice and Clio, and Clio and Bob. This opens the door to time travel paradoxes in the classical domain. Ralph  \cite{Ralph2011} first pointed this out for P-CTCs, but Ralph's procedure for a `radio to the past' is flawed, as we show below.

As a further consequence, since both D-CTCs and P-CTCs allow classical information to be sent around a spacetime loop, it follows from the Aaronson and Watrous result about CTC-enhanced \emph{classical} computation that a quantum computer with access to P-CTCs would also have the power of PSPACE, equivalent to quantum computers with access to D-CTCs.

 \section{D-CTCs}
 
Deutsch's consistency condition (\ref{D-consis}) applies to quantum CTC interactions $\rho_{CTC}\otimes\rho_{CR} \rightarrow U(\rho_{CTC}\otimes\rho_{CR} )U^{\dagger}$, where $\rho_{CR}$ is the state of the causality respecting (CR) system and $\rho_{CTC}$ is the state of the CTC system.

\begin{figure}[htb]
\begin{center}
\includegraphics*[width=8cm]{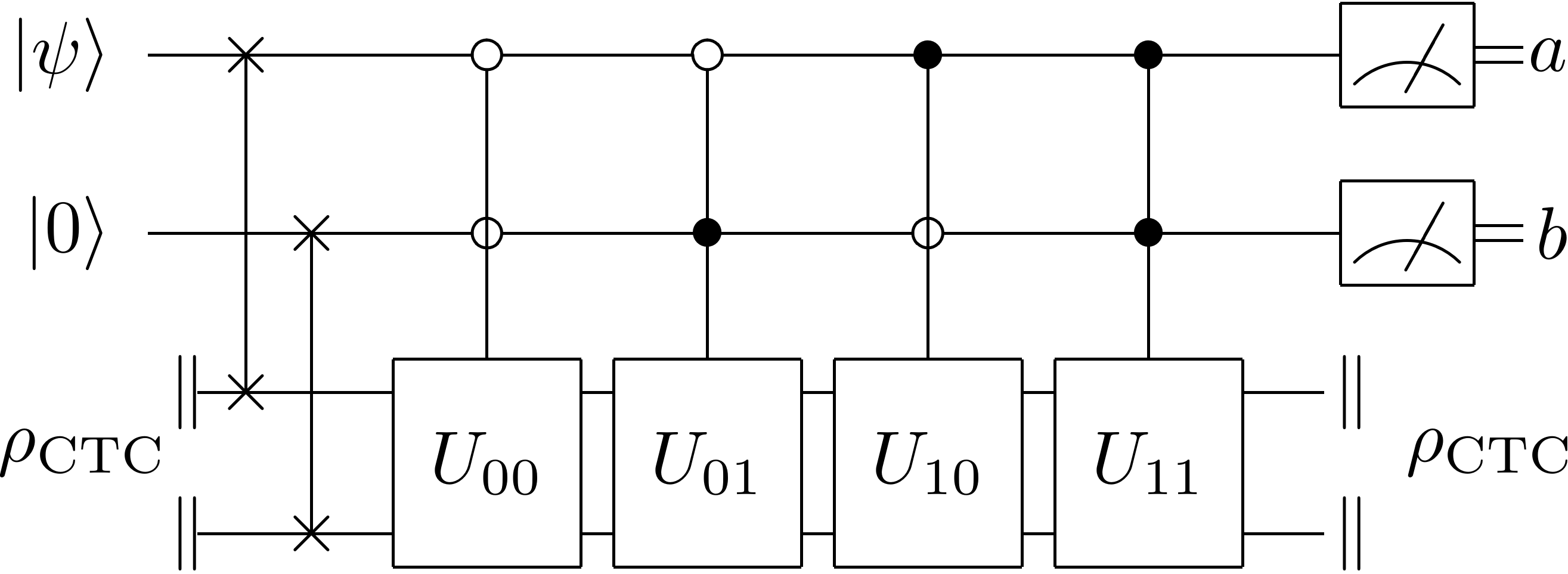}
\caption{A circuit that can perfectly distinguish the BB84 states $\ket{0}, \ket{1}, \ket{+}, \ket{-}$. From \cite{Brun+2009}.}
    \label{Brunfig}
    \end{center}
\end{figure}

The  Brun-Harrington-Wilde (BHW) circuit  \cite{Brun+2009} (see Figure \ref{Brunfig}) distinguishes between the BB84 input states $|0\rangle, |1\rangle, |+\rangle, |-\rangle$ for $|\psi\rangle$, where $\ket{\pm} = \frac{1}{\sqrt{2}}(\ket{0} \pm \ket{1})$. The unitaries are: $U_{00} = SWAP, U_{01} = X \otimes X, U_{10} = (X\otimes I)\circ (H \otimes I),  U_{11} = (X \otimes H) \circ (SWAP)$. The  `$\circ$' denotes composition of the operations, in order. The circuit notation is the standard notation for quantum circuits (see, e.g., \cite{NielsenChuang}). Reading from left to right, the crosses swap the states of the two qubits. The inputs to the black and white circles are control qubits for the unitary in the box, which is applied to the target qubit  if and only if the input to a black circle is $\ket{1}$ and the input to a white circle is $\ket{0}$. The boxes with $a$ and $b$ outputs represent measuring instruments with outcomes $a$ and $b$ in the computational basis. The unknown state here is $\ket{\psi}$; the state $\ket{0}$ is an ancilla state. The circuit uses two CTC qubits. The double vertical bars on the bottom left and right indicate the past and future mouths of the CTC wormhole.  

The circuit implements the following map on the unknown input state and the ancilla state, in order:
\begin{equation}
\begin{alignedat}{2}
\ket{0}\ket{0} &\rightarrow  \ket{0}_{a}\ket{0}_{b},\;\; \ket{+}\ket{0}&&\rightarrow \ket{1}_{a}\ket{0}_{b} \\
\ket{1}\ket{0} &\rightarrow  \ket{0}_{a}\ket{1}_{b},\;\; \ket{-}\ket{0}&&\rightarrow \ket{1}_{a}\ket{1}_{b}
\end{alignedat}
\end{equation}
where subscripts `$a$' and `$b$' on the output states indicate the states received by the two measuring instruments. So if $a = 0$,  $\ket{\psi}$ is a computational basis state, either $\ket{0}$ if $b=0$, or $\ket{1}$ if $b=1$. If $a=1$, $\ket{\psi}$ is one of the states $\{\ket{+},\ket{-}\}$, depending on the value of $b$.

The fixed point solutions (in this case unique) for $\rho_{CTC}$ are
\begin{eqnarray}
\ket{\psi} =   \ket{0} & \Longrightarrow & \rho_{CTC}= \ket{0}\ket{0} \\
\ket{\psi}  =   \ket{1} &\Longrightarrow &\rho_{CTC}= \ket{0}\ket{1} \\
\ket{\psi}  =   \ket{+} & \Longrightarrow & \rho_{CTC}=\ket{1}\ket{0} \\
\ket{\psi}  =  \ket{-} & \Longrightarrow & \rho_{CTC}= \ket{1}\ket{1}
\end{eqnarray}
To see this, take the input state $\ket{+}$ as an example. The swaps marked by crosses replace the states on the first two lines of the circuit with the CTC state $\ket{1}\ket{0}$, so only the unitary $U_{10}$ is applied to the input state $\ket{+}$ and ancilla state $\ket{0}$, now on the bottom two lines. The effect is to reproduce the initial CTC state: $U_{10}\ket{+}\ket{0} = \ket{1}\ket{0}$ so that Deutsch's consistency condition (\ref{D-consis})  is satisfied.

If the input state is the maximally mixed state $I/2$, the fixed point solution is
\begin{equation}
\rho_{CTC}= I/2 \otimes I/2 
\end{equation}
To see this, notice that after the SWAPS:
\begin{eqnarray}
\rho_{CR} & = &  I/2 \otimes I/2  \\
\rho_{CTC} & =  & I/2 \otimes |0\rangle\langle 0|
\end{eqnarray}
Each of the four unitaries is activated with probability 1/4:
\begin{eqnarray}
U_{00}: I/2\otimes |0\rangle \langle 0| & \longrightarrow & |0\rangle\langle0|\otimes I/2 \\
U_{01}: I/2\otimes |0\rangle \langle 0| & \longrightarrow & I/2 \otimes |1\rangle\langle 1| \\
U_{10}: I/2\otimes |0\rangle \langle 0|  & \longrightarrow & I/2 \otimes |0\rangle\langle 0| \\
U_{11}: I/2\otimes |0\rangle \langle 0| & \longrightarrow & |1\rangle\langle 1|\otimes I/2 
\end{eqnarray}
So the final CTC state exiting the circuit is
\begin{multline}
1/4(|0\rangle\langle0|\otimes I/2 +  I/2 \otimes |1\rangle\langle 1| + I/2 \otimes |0\rangle\langle 0| + |1\rangle\langle 1|\otimes I/2) \\=
 I/2 \otimes I/2
\end{multline}

Suppose Alice and Bob are spacelike separated and share a Bell state: 
\begin{equation}
\frac{1}{\sqrt{2}}(|0\rangle_{A}|1\rangle_{B} + |1\rangle_{A}|0\rangle_{B}) = \frac{1}{\sqrt{2}}(|+\rangle_{A}|+\rangle_{B} - |-\rangle_{A}|-\rangle_{B})
\end{equation}
Alice measures her qubit  in the computational basis or the diagonal basis, and Bob feeds his qubit into the BHW circuit.  In a reference frame  in which Alice measures in the computational basis or the diagonal basis \emph{before} Bob feeds his qubit into the circuit, Alice disentangles the Bell state so that Bob feeds a pure state, $|0\rangle$ or $|1\rangle$ into the circuit,  or $|+\rangle$ or  $|-\rangle$ into the circuit.
In a frame in which Alice measures \emph{after} Bob feeds his qubit into the circuit, Bob feeds the mixed state $I/2$ into the circuit. 

There is now a question of consistency between the two cases:
\begin{itemize}
\item Observers $O_{1}$: Alice measures her qubit before Bob feeds his qubit into the circuit.
If Alice measures in the computational basis, Bob feeds a pure state $|0\rangle$ or $|1\rangle$ into the circuit, and  \emph{the register $a$ reads 0  with probability 1}.
If Alice measures in the diagonal basis, Bob feeds a pure state $|+\rangle$ or $|-\rangle$ into the circuit, and \emph{the register $a$ reads 1 with probability 1}.
\item Observers $O_{2}$: Alice measures her qubit after Bob feeds his qubit into the circuit.
Bob feeds the mixed state $I/2$ into the circuit and \emph{the register $a$ reads 0 or 1 with probability 1/2}.
\end{itemize}

We propose the following \emph{consistency condition} to ensure consistency between observers $O_{1}$ and $O_{2}$:
\begin{itemize}
\item[(C1)]  Observers in differently moving reference frames agree on which events occur, even if they disagree about the order of events.
\item[(C2)] If an event has zero probability in any frame of reference, it does not occur.
\end{itemize}

For observers $O_{2}$, events occur with probability 1/2 that have zero probability for observers $O_{1}$ (register $a$ reads 0 and Alice measures in the diagonal basis, or register $a$ reads 1 and Alice measures in the computational basis). 
We achieve consistency between all frames only if we give priority to the description of events from the perspective of observers $O_{1}$ for whom Alice's measurement disentangles the state before Bob feeds his qubit into the circuit. It follows that the possibility of distinguishing between computational basis states and diagonal basis states enables effectively   instantaneous signaling in this spacelike separated case, even though nothing physical passes between Alice and Bob.

Without the consistency condition one might conclude that D-CTCs are inconsistent, as Lobo \emph{et al.} do \cite{Lobo+2011}:
\begin{quote}
How can this be possible? The only conceivable answer is that it \emph{cannot} be. We summarize our analysis in the following way: In one reference system, Bob's state is pure and he manages to receive superluminal communication which ultimately leads to those same paradoxes that Deutsch was originally trying to prevent and so it is then inconsistent. In another reference frame, his state is an improper mixture and the signaling protocol fails. But this is inconsistent with the first scenario. We conclude then that Deutsch's CTC model itself is \emph{overall} inconsistent. 
\end{quote}
Our consistency condition avoids this dire conclusion.

It is possible for Alice and Clio to be spacelike separated, and Clio and Bob to be spacelike separated, so that Bob is in Alice's causal past. 
Suppose
Alice and Clio share a Bell state
\begin{equation}
|\phi^{+}\rangle_{AC_{1}} = \frac{1}{\sqrt{2}}(|0\rangle_{A} |1\rangle_{C_{1}} + |1\rangle_{A}|0\rangle_{C_{1}})
\end{equation}
and Clio and Bob share a Bell state 
\begin{equation}
|\phi^{+}\rangle_{C_{2}B} = \frac{1}{\sqrt{2}}(|0\rangle_{C_{2}} |1\rangle_{B} + |1\rangle_{C_{2}}|0\rangle_{B})
\end{equation}
If Alice measures her qubit $A$ either in the computational basis or the diagonal basis, and Clio has access to a BHW circuit, then Alice can signal to Clio. If Clio measures her qubit $C_{2}$ in the computational basis or the diagonal basis, depending on the signal she receives from Alice, and Bob has access to a separate BHW circuit, then Clio can relay Alice's signal to Bob.

We conclude that signaling together with relaying yields a procedure for signaling to the past. D-CTCs were designed to ensure the existence of a fixed point solution that evades classical time travel paradoxes. Assuming that measurements have definite outcomes, relaying allows  classical information to be sent around a spacetime  loop and used  to trigger classical devices, opening the door to paradoxes in the classical domain.   For example, a D-CTC relay circuit could mimic any classical grandfather paradox scenario. We know in advance that we won't kill Grandpa---logic assures us of that---but D-CTCs don't provide any systematic explanation of why we fail. 

Deutsch would presumably resist this conclusion. His consistency condition comes with a caveat \cite[p. 3206]{Deutsch1991}:
\begin{quote}
Now recall the consistency condition for the evolution round a closed timelike line. In the quantum case I have taken it to be that the density operator of each chronology-violating bit must return to its original value at a given event, as expressed by (15). That is the correct condition under the unmodified quantum formalism, \emph{but it is either wrong or insufficient under every other version of quantum theory}, just as under classical physics.
\end{quote}
The reference to equation (15) here is to our equation (\ref{D-consis}). The `unmodified quantum formalism' is Deutsch's term for the Everett interpretation of quantum mechanics---he regards other interpretations as modified versions of the theory. On Deutsch's view \cite[p. 3207]{Deutsch1991}: `Closed timelike lines would provide gateways between Everett universes.' 

We do not take a position here on the issue of interpretation. There is a rich literature of results about the information-theoretic power of D-CTC interactions, following Deutsch's seminal paper,  and none of this hinges on anything more than operational features of quantum mechanics. Here we take note of these results and demonstrate some further unexpected consequences. Perhaps more to the point, our consistency condition shows how Deutsch's model of quantum CTC interactions can avoid the sort of inconsistency pointed out by Lobo \emph{et al.} \cite{Lobo+2011}  without presupposing an Everettian interpretation.

\begin{figure}[htb]
\begin{center}
\includegraphics*[width=3.5cm]{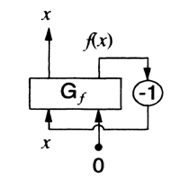}
\caption{Network for finding a fixed point of $f$. From \cite[Fig. 5]{Deutsch1991}.}
    \label{Deutschfig}
    \end{center}
\end{figure}

We note that Deutsch is particularly concerned to show that his fixed point consistency condition excludes the possibility of an unproved theorem paradox: Bob reads the proof of a theorem in a journal and sends it back to Alice in the past, who publishes the proof in the journal that Bob subsequently reads. The `paradox' is that the theorem comes from nowhere, without any intellectual effort from Alice or Bob. Deutsch regards this (`paradox 4') as `a far more serious paradox' than the grandfather paradox \cite[p. 3202]{Deutsch1991}. 
The unproved theorem paradox is associated with the network in Figure \ref{Deutschfig} for finding the fixed point of a function $f$ in one step (where the $-1$ represent a negative temporal increment, i.e., time travel to the past). Our relay procedure shows that precisely such a circuit can be achieved with  D-CTCs. 

Jacques Pienaar \footnote{Private communication.} has pointed out that a variant of quantum theory  proposed by Adrian Kent \cite{Kent2005a,Kent2005b} would block signaling to the past via relayed spacelike separated scenarios involving D-CTCs. For an entangled state of $n$ qubits and local projective measurements on some of the $n$ qubits, the theory suggests a novel relativistically invariant definition of the `local state' of a qubit. In the case of a Bell state of two qubits, where one of the qubits is measured in the computational basis, the local state of the remote qubit is the random mixture $I/2$, until a sufficient amount of time has elapsed to causally connect the measurement event with the remote qubit, after which the local state is the appropriate pure state.  In considering the interaction of a CR qubit with a CTC qubit when the CR qubit is part of an entangled pair, the local state of the CR qubit would then be the mixture $I/2$  until  a measurement on the paired qubit could be causally connected  with the CR qubit. In that case, there would be no inconsistency between different reference frames for spacelike separated events, so our consistency condition would not be applicable and there would be no signaling to the past via relayed spacelike separated scenarios. A theory of this sort is an interesting speculation as a possible way of avoiding our conclusion---until it is excluded by experiment (see \cite{Cavalcanti+2012}).

\section{P-CTCs}

Brun and Wilde \cite{Brun+2012} show how it is possible with a P-CTC interaction to postselect or `force' any desired outcome of a measurement, except for a zero probability  outcome orthogonal to the state.

Begin by applying the measurement unitary to the state $\ket{\psi}= c_{1}\ket{0} + c_{2}\ket{1}$ and an ancilla in state $\ket{0}$ (the `pointer' state):
\begin{equation}
\ket{\psi}\ket{0} \stackrel{U}{\longrightarrow} c_{1}\ket{0}\ket{0} + c_{2}\ket{1}\ket{1}
\end{equation}
To `force' the measurement outcome $0$, apply $U_{0} = \ket{0}\bra{0}\otimes I + \ket{1}\bra{1}\otimes X$ to the ancilla state and the state $\ket{\;\,}_{1}$  of the Bell state $\frac{1}{\sqrt{2}}(\ket{0}_{1}\ket{0}_{2} + \ket{1}_{1}\ket{1}_{2})$ that drives the P-CTC interaction.
The final state is 
\begin{multline} 
c_{1}\ket{0}\ket{0}\otimes \frac{1}{\sqrt{2}}(\ket{0}_{1}\ket{0}_{2} + \ket{1}_{1}\ket{1}_{2}) \\+ c_{2}\ket{1}\ket{1}\frac{1}{\sqrt{2}}(\ket{1}_{1}\ket{0}_{2} + \ket{0}_{1}\ket{1}_{2})
\end{multline}
To simulate the P-CTC interaction, project qubits 1 and 2 onto the Bell state and renormalize.
The final CR state is $\ket{0}\ket{0}$. To `force' the measurement outcome $1$, apply $U_{1} = \ket{0}\bra{0}\otimes X + \ket{1}\bra{1}\otimes I$.
The final CR state is $\ket{1}\ket{1}$.

Evidently, if Alice and Bob share a Bell state:
\begin{equation}  
\frac{1}{\sqrt{2}}(|0\rangle_{A}|0\rangle_{B} + |1\rangle_{A}|1\rangle_{B})
\label{Bellstate}
\end{equation}
Alice can signal to Bob using a Brun-Wilde P-CTC circuit to `force' the outcome $0$ or $1$ of a computational basis measurement on her qubit $A$. Bob receives the signal by measuring his qubit $B$ in the computational basis. As in the discussion of D-CTCs in the previous section, the consistency condition applied to spacelike separated scenarios reconciles the different descriptions of  (i) reference frames in which Alice applies the unitary transformation $U_{0}$ or $U_{1}$ to her qubit $A$ and initiates the interaction with the CTC qubit (simulated by projection and renormalization) \emph{before} Bob measures his qubit $B$ (so that Alice applies $U_{0}$ or $U_{1}$ to an entangled state), and (ii) reference frames in which Alice applies the unitary and initiates the CTC interaction \emph{after} Bob's measurement  (so that Alice applies the unitary to a product state $|0\rangle_{A}|0\rangle_{B}$ or $|1\rangle_{A}|1\rangle_{B}$). Frames in which Bob measures first permit events such as a $0$ result for Bob, together with a failed attempt by Alice to force her measurement to have result $1$. Such events have probability zero in frames in which Alice applies $U_{0}$ or $U_{1}$ and allows her qubit to interact with the CTC qubit before Bob's measurement. The detailed analysis is similar to the analysis that follows for Ralph's circuit, showing that consistency requires giving priority to the description provided by frames in which Alice applies $U_{0}$ or $U_{1}$ to the entangled state $\frac{1}{\sqrt{2}}(|0\rangle_{A}|0\rangle_{B} + |1\rangle_{A}|1\rangle_{B})$. Once again, signaling to the past is possible by relaying spacelike separated signaling scenarios.

Ralph  \cite{Ralph2011} first proposed a P-CTC circuit for signaling to the past. Ralph's argument is flawed, but  it is illuminating to consider Ralph's circuit in some detail rather than the Brun-Wilde circuit to clarify the role of the consistency condition in P-CTC interactions. The flaw, as we see it, raises an important question of interpretation for P-CTCs.

\begin{figure}[htb]
\begin{center}
\includegraphics*[width=9.5cm]{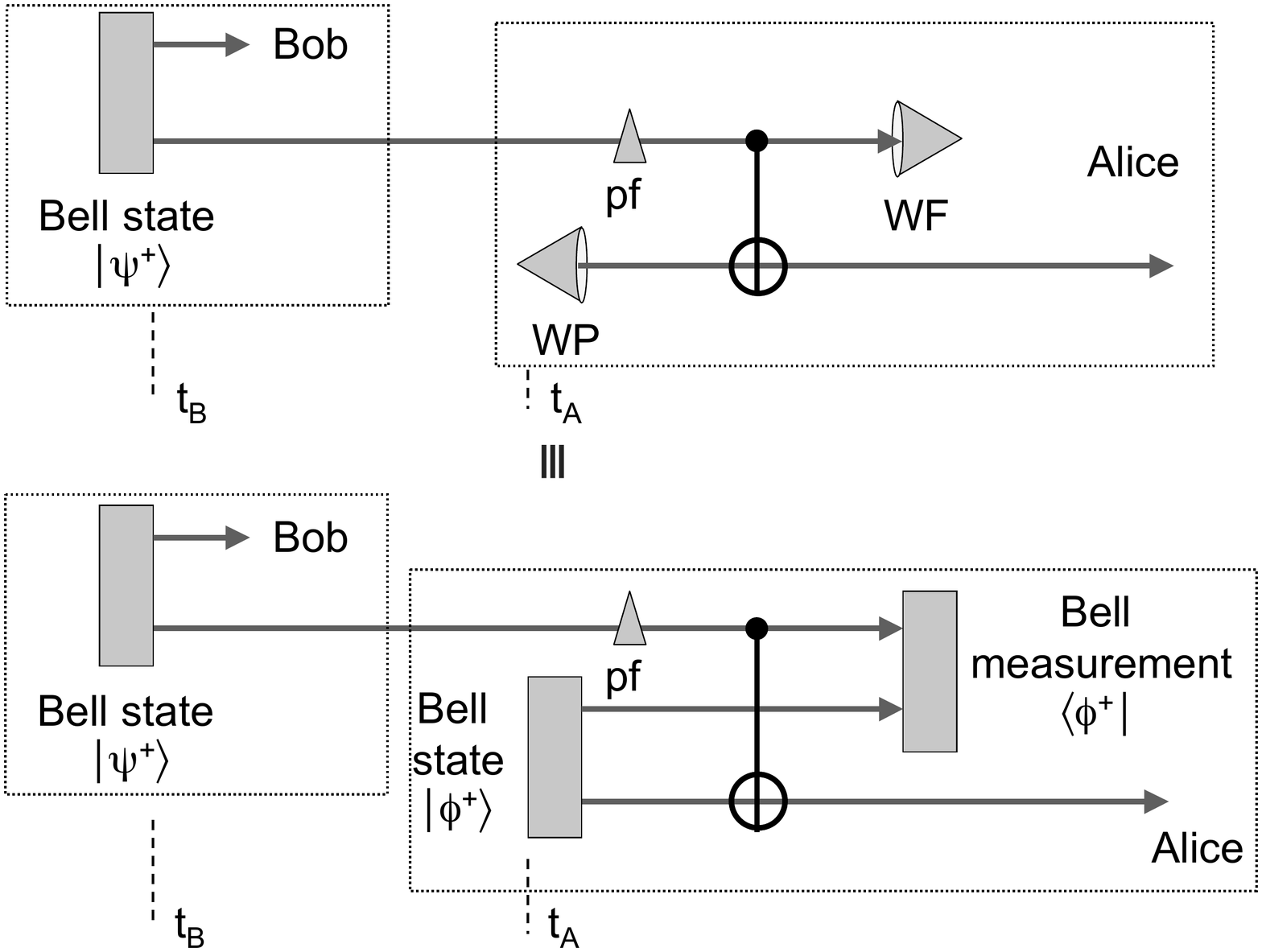}
\caption{Ralph's `radio to the past' circuit,' adapted from \cite{Ralph2011}. Top panel: model of entanglement interacting with a closed timelike curve formed by a wormhole. WF is the future mouth of the wormhole; WP is the past mouth of the wormhole; pf is a phase flip that may be applied. The interaction is a CNOT gate. Bottom panel: according to \cite{Lloyd+2011a}  this is equivalent to the depicted post-selected teleportation circuit in which the input state is renormalized to the post-selected outcome of the Bell measurement.}
  \label{Ralphfig}
\end{center}
\end{figure}

In Ralph's procedure for signaling to the past, Bob prepares the initial global state as the Bell state
\begin{eqnarray}
 |\psi^{+}\rangle & = & \frac{1}{\sqrt{2}}(|0\rangle_{B}|1\rangle_{A} + |1\rangle_{B}|0\rangle_{A}) \nonumber \\
 & = &  \frac{1}{\sqrt{2}}(|+\rangle_{B}|+\rangle_{A} - |-\rangle_{B}|-\rangle_{A})
 \end{eqnarray}
  at time $t_{B}$. He keeps one qubit and sends the other qubit to Alice. At a later time $t_{A}$, Alice creates a compact wormhole time machine by exploiting CTCs that form for times $t > t_{A}$. She chooses to either perform a phase flip on her qubit or not before passing the qubit  into the future mouth of a wormhole. After traveling a short time to the past, Alice's qubit emerges from the past mouth of the wormhole at $t_{A}$ and interacts with its past self via a CNOT gate. See Figure \ref{Ralphfig}, top panel. (To preserve consistency with our discussion of signaling via the BHW circuit and via the Brun and Wilde procedure for `forcing' a particular measurement outcome, in which Alice signals to Bob, we have modified Ralph's original diagram so that Alice signals to Bob rather than Bob to Alice. We have also replaced Ralph's labels for Bell states with the standard labels: $\ket{\psi^{+}} = \frac{1}{\sqrt{2}}(\ket{0}\ket{1}+\ket{1}\ket{0}), \ket{\phi^{+}} = \frac{1}{\sqrt{2}}(\ket{0}\ket{0}+\ket{1}\ket{1})$.) Following the prescription in \cite{Lloyd+2011a},  Ralph obtains the output of the interaction  from the equivalent teleportation circuit in the bottom panel of Figure \ref{Ralphfig} by projecting and  renormalizing to the outcome for which the same Bell state is detected at the Bell measurement as the Bell state that drives the teleportation. 

The initial total state is
\begin{equation}
|\psi^{+}\rangle\otimes |\phi^{+}\rangle = \frac{1}{2}(|0\rangle_{B}|1\rangle_{A} +  |1\rangle_{B}|0\rangle_{A})\otimes(|0\rangle_{1}|0\rangle_{2} + |1\rangle_{1}|1\rangle_{2})
\end{equation}
where 1 and 2 label the qubits in the entangled state in the teleportation circuit in the bottom half of Figure \ref{Ralphfig}. Alice either applies a phase flip to $A$, which induces the transformation $|0\rangle_{A} \rightarrow |0\rangle_{A}, |1\rangle_{A} \rightarrow -|1\rangle_{A}$, and a C-NOT gate to $A$ and qubit $1$  of the entangled state (with $A$ as control), or she does nothing to $A$ and applies a C-NOT gate to $A$ and qubit $1$.  Depending on whether there is a phase flip or not, the state after the C-NOT gate is equivalent to: 
\begin{multline}
\frac{1}{2}|0\rangle_{B}|1\rangle_{A}\otimes(|0\rangle_{1}|1\rangle_{2}+ |1\rangle_{1}|0\rangle_{2}) \\  \pm \frac{1}{2}|1\rangle_{B}|0\rangle_{A}\otimes(|0\rangle_{1}|0\rangle_{2}+ |1\rangle_{1}|1\rangle_{2}))
\end{multline}
where the $-$ sign corresponds to the phase flip case.
The causality respecting state after the CTC qubit enters the wormhole is obtained by projecting $|\hspace{.1in}\rangle_{A}|\hspace{.1in}\rangle_{1}$ onto $|\phi^{+}\rangle$ and renormalizing. This leaves the causality respecting state as
\begin{itemize}
\item no phase flip: $(|0\rangle_{B} + |1\rangle_{B})|0\rangle_{2} = |+\rangle_{B}|0\rangle_{2}$
\item phase flip: $(|0\rangle_{B} - |1\rangle_{B})|0\rangle_{2} = |-\rangle_{B}|0\rangle_{2}$
\end{itemize}
So Bob's qubit $B$ is left in the state $|+\rangle_{B}$ or $|-\rangle_{B}$ depending on whether or not Alice applies a phase flip to her qubit $A$.

Ralph concludes his argument with the comment:
\begin{quote}
\noindent By choosing to introduce a phase flip on [her] qubit or not, [Alice] can deterministically send a string of bits (in the diagonal basis) to [Bob] in the past. The existence of this `radio to the past' creates the possibility of new paradoxes, now in the classical domain. In particular, the unproved theorem paradox can arise.'
\end{quote}

Now, \textit{prima facie} Ralph's circuit  cannot work as proposed because Bob would have to measure his qubit in the diagonal basis to extract any information, and this measurement would have to be in Alice's backward light cone. 
In all inertial frames, Bob's measurement disentangles the Bell state, so the state Alice receives is either $|+\rangle$ or $|-\rangle$.
The notion of a radio to the past assumes that  Alice is free to choose whether or not to apply the phase flip. 
So it must be possible for Alice to receive $|+\rangle$ and apply the phase flip, or to receive $|-\rangle$ and not apply the phase flip. 
In each case, the total state after the C-NOT gate is orthogonal to the postselection subspace, so the projection is null.

If Bob's measurement outcome is $+$, the  initial state is
\begin{equation}
|+\rangle_{B}| + \rangle_{A} \otimes |\phi^{+}\rangle
\end{equation}
After 
the phase flip and C-NOT, the state is 
\begin{multline}
\frac{1}{2}|+ \rangle_{B}|0\rangle_{A}(|0\rangle_{1}|0\rangle_{2}+ |1\rangle_{1}|1\rangle_{2}) \\ - \frac{1}{2}|+\rangle_{B}|1\rangle_{A}(|0\rangle_{1}|1\rangle_{2}+ |1\rangle_{1}|0\rangle_{2})
\end{multline}

If Bob's measurement outcome is $-$, the initial state is
\begin{equation}
|-\rangle_{B}|-\rangle_{A} \otimes |\phi^{+}\rangle
\end{equation} 
After no phase flip and C-NOT, the state is
\begin{multline}
\frac{1}{2}|- \rangle_{B}|0\rangle_{A}(|0\rangle_{1}|0\rangle_{2} +  |1\rangle_{1}|1\rangle_{2}) \\ - \frac{1}{2}|-\rangle_{B}|1\rangle_{A}(|0\rangle_{1}|1\rangle_{2}+ |1\rangle_{1}|0\rangle_{2})
\end{multline}

In both of these cases, the final state is orthogonal to the postselection subspace, so the projection is null.

On the standard interpretation of P-CTCs (\cite{Lloyd+2011a}, \cite{Ralph2011}),   interactions with P-CTCs produce the required postselection \textit{by stipulation}. Since Alice's qubit will interact with the CTC qubit, cases in which Alice applies a phase flip to $|+\rangle$ or does not apply a phase flip to $|-\rangle$ are ruled out. 
This would mean that if Alice applies a phase flip then this \textit{causes} Bob's measurement to have the outcome $|-\rangle$, and if she does not, this \textit{causes} Bob's measurement to have the outcome $|+\rangle$.

It seems just as plausible to say that if the state on which the C-NOT (or other) gate would act conflicts with the postselected state, the interaction fails to occur.
This avoids a gratuitous constraint on events in the past. Note that D-CTCs avoid this: \textit{any} state can interact with a D-CTC.
But on this interpretation, Ralph's circuit does not work as a radio to the past.
When Bob measures $+$, he knows only that either Alice will not apply the phase flip, or the CTC interaction will not occur, and when he measures $-$ he knows only that either Alice will apply the phase flip, or the CTC interaction will not occur.

The consistency condition allows one to  sidestep this issue. 
Suppose Bob's measurement and Alice's action (phase flip or not, followed by the C-NOT gate) are  spacelike separated, so there are frames in which Bob measures before Alice's action and disentangles the Bell state, and frames in which Alice acts before Bob's measurement on the entangled state. 

Only the entangled state perspective respects the consistency condition.
If Alice applies the unitary transformations (phase flip or identity, followed by C-NOT) to the entangled state, then we always have a non-null projection onto the post-selected state. This means, in terms of an interaction with the CTC qubit, that the CTC interaction always occurs.
If Alice applies the unitary transformations to one component of the entangled state (corresponding to Bob's measurement outcome), then we do or we do not have a non-null projection onto the post-selected state, depending on the component (i.e., on Bob's measurement outcome) and on whether Alice applies the phase shift or not (because in order to get a non-null projection in each case, we need both components of the entangled state).

For an observer $O_{1}$ for whom Bob's measurement occurs first, Alice applies the unitaries 
to her part of a product state. For $O_{1}$ each of the following is possible:
\begin{itemize}
\item Bob's measurement yields $+$ and Alice doesn't apply the flip
\item Bob's measurement yields $-$ and Alice applies the flip
\item Bob's measurement yields $+$ and Alice applies the flip
\item Bob's measurement yields $-$ and Alice doesn't apply the flip
\end{itemize}
For an observer $O_{2}$ for whom Alice's action occurs first, Alice applies the unitaries to her part of an entangled state. This rules out the last two possibilities, which have zero probability for $O_{2}$. For $O_{2}$ the interaction with the CTC qubit always occurs (the projection is always non-null). 
Consistency between observers requires that this must be the case for $O_{1}$ as well.

So \emph{however we interpret the significance of a null projection}, superluminal signaling with P-CTCs is possible, and given this possibility, a relay procedure allows signaling to the past. Suppose Alice and Clio are spacelike separated, and Clio and Bob are spacelike separated, so that Bob is in Alice's causal past. Suppose
Alice and Clio share a Bell state
\begin{equation}
|\psi^{+}\rangle_{C_{1}A} = \frac{1}{\sqrt{2}}(|0\rangle_{C_{1}} |1\rangle_{A} + |1\rangle_{C_{1}}|0\rangle_{A})
\end{equation}
and Clio and Bob share a Bell state 
\begin{equation}
|\psi^{+}\rangle_{BC_{2}} = \frac{1}{\sqrt{2}}(|0\rangle_{B}|1\rangle_{C_{2}} + |1\rangle_{B}|0\rangle_{C_{2}})
\end{equation}
Alice feeds her qubit $A$ 
 into a Ralph circuit and either flips the phase or not to send a bit to Clio.
Clio measures her qubit $C_{1}$ in the diagonal basis to receive Alice's bit.
Depending on the bit she receives, Clio either flips the phase of her qubit $C_{2}$ or not.
Bob measures his qubit $B$ in the diagonal basis to receive Clio's bit.
Then Alice's action, phase flip or not, is correlated with Bob's measurement outcome in Alice's 
causal past.

Lloyd \emph{et al.} \cite {Lloyd+2011b} argue that Ralph does not succeed in generating an unproved theorem paradox, as he claims. 
They pose a dilemma: either one of Alice or Bob is the author of the theorem, or  not. If one of them is the author, there is no paradox: the theorem doesn't simply `appear.'
For the theorem simply to `appear,'  it cannot depend on what Alice or Bob \textit{choose} to do. So, to avoid the second horn of the dilemma, Alice's phase flip must be part of a control gate, where the control is ultimately provided by Bob's measurement outcome. 
Lloyd \emph{et al.} provide no general proof that such a circuit is impossible. They merely sketch two examples in which things go wrong. In one case, the proposal is inconsistent with the postselection occurring. In the second case, the `theorem' turns out to be trivial.

Even if there were a circuit similar to Ralph's that could produce an unproved theorem, with no hint that Alice or Bob is the author, this would not settle the dispute in Ralph's favor. The fact that a model of CTCs \emph{permits} unproved theorems to appear from nowhere is not a genuine problem.
Alice could decide whether or not to apply a phase flip in Ralph's circuit by flipping a coin. The result \emph{could} be a theorem, but Alice would be the author only in a Pickwickian sense. It is \emph{always} possible that a sequence of binary outcomes will encode a theorem merely by chance as a  very low-probability event, whether or not a `radio to the past' is part of the process. In short, if there were a `radio to the past,' it \emph{could} produce an output that encodes a theorem, but this would only  be problematic if a model for CTCs made it \emph{likely} that a non-trivial unproved theorem would appear.

\section{Conclusion}

Deutsch's model of quantum CTC interactions was designed to avoid imposing gratuitous constraints to  thwart time travel paradoxes like the grandfather paradox or the unproved theorem paradox. Logic tells us that such paradoxical situations cannot occur, but in a classical theory the only explanation available for the non-occurrence of paradox is a `banana peel' explanation: something happens, like slipping on a banana peel that just  happens to be conveniently placed so that the paradoxical event  fails to occur. P-CTCs are supposed to avoid paradox because anything that \textit{could} happen in a P-CTC circuit \textit{does} happen with some probability in an ordinary teleportation circuit. As Lloyd \emph{et al.} \cite[p. 025007]{Lloyd+2011a} emphasize: `Because the theory of P-CTCs rely on post-selection, they provide self-consistent resolutions to such paradoxes: anything that happens in a P-CTC can also happen in conventional quantum mechanics with some probability.' 

We have shown that both P-CTCs and D-CTCs allow the possibility of a `radio to the past' that operates in the classical domain, in which there is no systematic fix to time-travel paradoxes. As a consequence, it follows that CTC-enhanced quantum computers have the power of PSPACE, for both D-CTC and P-CTC quantum interactions.

This follows from our consistency condition and relay procedure.  Aaronson and Watrous \cite{Aaronson+2009}  proved that classical computers with access to CTCs would have the power of PSPACE, equivalent to the power of  D-CTC-enhanced quantum computers. Since our relay procedure with P-CTC interactions allows classical information to be sent around a spacetime loop and used to trigger classical devices, it follows that P-CTC-enhanced quantum computers would also have the power of PSPACE, rather than (merely) PP. Equivalently, this conclusion follows from the ability to construct the fixed point network of Figure \ref{Deutschfig}.

\section*{Acknowledgements} We thank Scott Aaronson, Jacques Pienaar, and Lorenzo Maccone for comments on a previous version of the paper. Jeffrey Bub's research is supported by the Institute for Physical Science and Technology at the University of Maryland, and his work on this publication was made possible through the support of a grant from the John Templeton Foundation. The opinions expressed in this publication are those of the author and do not necessarily reflect the views of the John Templeton Foundation.

\bibliographystyle{apsrev}
\bibliography{CTC.bib}

\end{document}